\documentclass[12pt,preprint]{aastex}
\usepackage{slashbox} 
\newcommand{\del}[2]%
{\frac{\mathrm{d}{#2}}{\mathrm{d}{#1}}}
\newcommand{\Del}[2]%
{\frac{\mathrm{D}{#2}}{\mathrm{D}{#1}}}
\newcommand{\ddel}[2]%
{\frac{\mathrm{d}^2{#2}}{\mathrm{d}{#1}^2}}

\newcommand{\pdel}[2]%
{\frac{\partial{#2}}{\partial{#1}}}
\newcommand{\pddel}[2]%
{\frac{\partial^2{#2}}{\partial{#1}^2}}

\newcommand{\laplace}{\bigtriangleup}

\renewcommand{\vec}[1]{\mathbf{#1}}

\newcommand{\Ms}{M_{\odot}}
\newcommand{\km}{\mathrm{km}}
\newcommand{\cm}{\mathrm{cm}}
\newcommand{\ms}{\mathrm{ms}}
\newcommand{\cmps}{\mathrm{cm/s}}
\newcommand{\MRI}{\mathrm{MRI}}
\newcommand{\WRAP}{\mathrm{WRAP}}
\newcommand{\gauss}{\mathrm{G}}
\newcommand{\erg}{\mathrm{erg}}
\newcommand{\psec}{\mathrm{s}^{-1}}
\newcommand{\gpcmc}{\mathrm{g/cm^3}}

\def\d{\mathrm{d}}

\def\Rnum#1{\uppercase\expandafter{\romannumeral #1}}
\def\rnum#1{\expandafter{\romannumeral #1}}

\begin{document}

\title{Magneto-driven Shock Waves in Core-Collapse Supernova}

\author{Tomoya Takiwaki\altaffilmark{1},
Kei Kotake\altaffilmark{1},
Shigehiro Nagataki\altaffilmark{2},
 and Katsuhiko Sato\altaffilmark{1,3}}

\affil{\altaffilmark{1}Department of Physics,
School of Science, the University of Tokyo, 7-3-1 Hongo,
Bunkyo-ku,Tokyo 113-0033, Japan}
\affil{\altaffilmark{2}
Yukawa Institute for Theoretical Physics, Kyoto University,
Kyoto 606-8502, Japan}
\affil{\altaffilmark{3}
Research Center for the Early Universe,
School of Science, the University of Tokyo,7-3-1 Hongo,
Bunkyo-ku, Tokyo 113-0033, Japan}
\email{takiwaki@utap.phys.s.u-tokyo.ac.jp}

\begin{abstract}
We perform a series of two-dimensional  
magnetohydrodynamic
simulations of the rotational core-collapse 
of a magnetized massive star.
We employ a realistic equation of state
and take into account the neutrino cooling
by the so-called leakage scheme.
In this study 
we systematically investigate
how the strong magnetic field and the rapid rotation
affect the propagation of the shock waves.
Our results show that 
in the case of the strong initial poloidal magnetic field,
the toroidal magnetic field
amplified by the differential rotation,
becomes strong enough to
generate a tightly collimated shock wave along the rotational axis.
On the other hand,
in the case of the weak  initial magnetic field,
although the differential rotation  amplifies
 toroidal magnetic field over the long rotational period,
the launched shock wave is weak and 
the shape of it becomes wider.
The former case is expected to be accompanied by the formation
of the so-called magnetar.
Our models with rapid rotation and 
strong magnetic field can create a nozzle
formed by the collimated shock wave.
This might be the analogous 
situation of the collapsar 
that is plausible for the central engine of the Gamma-Ray Bursts.
\end{abstract}


\section{Introduction}

The study of collapse driven supernovae is 
important not only for itself  but also for the 
understanding of other process of astrophysical relevance,
such as nucleosynthesis of heavy elements and chemical 
evolutions in the universe \citep{arne96},
radiation of neutrinos in the universe \citep{Raff02},
the gravitational waves \citep{ande03},
and possibly gamma-ray bursts (GRB) and hypernovae \citep{macf99}.
In spite of its importance and extensive
investigations done so far,
the explosion mechanism has not been clarified yet.

Except for the special cases, the shock wave generated
by the core bounce stalls and becomes an accreting shock
in the core.
Although neutrino heating are expected to
revive the shock wave and lead to the successful
explosion \citep{wils85},
recent theoretical studies that elaborate
the neutrino transport method and detailed 
microphysics and/or general relativity,
failed to produce explosions \citep{ramp00,lieb01,thom03,bura03}.
However, it should be noted that most of
them assume spherical symmetry (see, however, Buras et al. 2003).

On the other hand, there are convincing
observations, which require the revision of 
the spherically symmetric stellar collapse.
Rather common correlation between the 
asymmetry and the collapse-driven supernovae 
has been reported by the observation of 
the polarization \citep{wang02}. 
It is also well known that SN 1987A is globally asymmetric
 \citep{crop88,papa89}, 
which is directly confirmed by the images of 
the {\it Hubble Space Telescope} \citep{plat95}.
The observed light curve also supports jet-like explosion \citep{naga97,naga00}. 
 The asymmetry is likely to have
originated from the core dynamics \citep{khok99}.
Provided the facts that the progenitors of the collapse-driven
supernovae are a rapid rotator on the main sequence \citep{tass78}
 and that the recent theoretical studies suggest
a fast rotating core prior to the collapse,
 it is important to incorporate rotations in
simulations of core collapse \citep{hege00}.

In addition, after the discovery of pulsars,
it was reasonable to explore the issue of 
whether or not the rotation and magnetic fields
associated with pulsars could be a significant factor
in the explosion mechanism.
So far there have been some works devoted 
to the understanding of the effect 
of rotation and magnetic field
on the supernova explosion mechanism 
\citep{lebl70,mull81,bode83,symb84,monc89,yama94,frye00,kota03b}.
The magnetocentrifugal jets generated 
by the strong toroidal magnetic fields 
in stellar collapse may explain why all core collapse
supernovae are found to be substantially asymmetric and 
predominantly bipolar \citep{whee02}. 
With the typical dipole fields of $10^{12}\gauss$ and 
rotation periods of several tens of milliseconds,
however implying electro dynamic power of only $\approx
10^{44}-10^{45}\erg\psec$,
a strong robust explosion seemed unlikely.
Several factors, such as observation of magnetar \citep{dunc92},
 may lead to a need to reexamine the conclusion.
For analyzing the origin of the strong magnetic field,
it is necessary to investigate the magnetohydrodynamic processes 
such as magnetorotational instability (MRI) and the field line 
wrapping mechanism \citep{ostr71,bisn71,bisn76,kund76,balb98,meir76,akiy03}.
These processes strongly depend on the configuration and the strength 
of the magnetic fields.
We think that it's still worth investigating the model of
strong magnetic field and rapid rotation 
those have not been considered sufficiently so far.
It should be mentioned that
 such models  may be related to some theoretical models of GRBs and 
magnetar \citep{prog03,blan03}.
By changing the strength of rotation
and the poloidal magnetic fields in a parametric manner,
we investigate how the rapid rotation and the strong 
magnetic fields affect the property of the shock waves
 and the explosion energy.
By so doing,
we hope to understand the difference
of the hydrodynamic behaviors 
between our models and the models for the canonical 
supernovae discussed so far.

We describe the numerical method and the initial model in the next section.
In the third section,
we show the numerical results.
The summary and some discussions are given in the last section.

\section{Method of Calculation}
\subsection{Basic Equations}
The numerical method for magnetohydrodynamic (MHD) computations employed in this paper
is based on the ZEUS-2D code \citep{ston92}.
The basic equations for the evolution are written as follows,
\begin{equation}
\Del{t}{\rho}+\rho\nabla\cdot\vec{v}=0,
\end{equation}
\begin{equation}
\rho \Del{t}{\vec{v}}=-\nabla P -\rho \nabla \Phi
+\frac{1}{4\pi}\left(\vec{\nabla}\times\vec{B}\right)\times\vec{B},
\end{equation}
\begin{equation}
\rho\Del{t}{}\left(\frac{e}{\rho}\right)=-P\nabla \cdot \vec{v}
-L_{\nu},
\end{equation}
\begin{equation}
\pdel{t}{\vec{B}}= \vec{\nabla} \times \left(\vec{v}\times			\vec{B}\right)\label{eq:1},
\end{equation}
\begin{equation}
\laplace{\Phi} = 4\pi G \rho,
\end{equation}
where $\rho,P,\vec{v},e,\Phi,\vec{B},L_{\nu}$ 
are the mass density,
the gas pressure
including the radiation pressure from neutrino's,
the fluid velocity, the internal energy density, the gravitational potential,
the magnetic field and the neutrino cooling rate, respectively.
We denote Lagrange derivative,
as $\Del{t}{}$.
The ZEUS-2D is an Eulerian code based on the finite-difference
method and employs an artificial viscosity of von Neumann and 
Richtmyer to capture shocks.
The time evolution of magnetic field is 
solved by induction equation, Eq. (\ref{eq:1}).
In so doing, the code utilizes the so-called
constrained transport (CT) method, which ensures
the divergence free($\nabla\cdot\vec{B}=0$) 
of the numerically evolved magnetic fields at all times.
Furthermore, the method of characteristics (MOC)
is implemented to propagate accurately all modes of MHD waves. 
The self-gravity is managed by solving the Poisson equation
with the incomplete Cholesky decomposition conjugate gradient
(ICCG) method.
In all the computations,spherical coordinates are used.
We made several major changes to the base code to 
include the microphysics.
First we added an equation for electron fraction
to treat electron captures and neutrino transport by the 
so-called leakage scheme \citep{epst81}.
The cooling rate, $L_{\nu}$, is also estimated by the scheme.
For a more detailed description of this scheme, see Kotake 2003 \citep{kota03a}.
Second we have incorporated the tabulated equation of state (EOS)
based on relativistic mean field theory instead of 
the ideal gas EOS assumed in the original code \citep{shen98}.   

\subsection{Initial Models}
We make precollapse models by taking the profile of density,
internal energy, electron fraction distribution from 
$20\Ms$ rotating presupernova model of Heger \citep{hege00},
and assume the profile of rotation and the magnetic field as follows.
As for the rotation profile, we assume the cylindrical rotation:
\begin{equation}
 \Omega(X,Z)=\Omega_{0}\frac{X_{0}^2}{X^2+X_{0}^2}\frac{Z_{0}^4}{Z^4+Z_{0}^4},
\end{equation}
where X and Z denote distance from rotational axis and the equatorial
plane.
We adopt the value of parameters, $X_0$ and $Z_0$, as  $10^7 \cm,10^8
\cm$, respectively.
We assume that the initial magnetic field is uniform and 
parallel to the rotational axis:
\begin{equation}
 B_{z}=B_{0}.
\end{equation}
Although recent stellar evolution studies show that 
toroidal magnetic field components may be stronger than
the poloidal ones prior to the collapse \citep{spru02},
uncertainty remains still in the model.
In this study we choose the  poloidal magnetic fields in order to see
 the mechanism of field amplification \citep{whee02}.
We compute 12 models changing
the total angular momentum and strength of magnetic field
by varying the value of $\Omega_0$ and $B_{0}$.
The model parameters are given in Table \ref{tab:model}.
The models are named 
after this combination,
with the first letters, B12, B10.5, B9, B0,
representing strength of initial magnetic field,
the following letter, TW1, TW2, TW4 
representing the initial $T/|W|$.
Here $T/|W|$ indicates the ratio of 
rotational energy to absolute value of gravitational energy.
\section{Numerical Results}

\subsection{The Hydrodynamic Features}\label{sec:hydr-feat}
\begin{figure}
\plotone{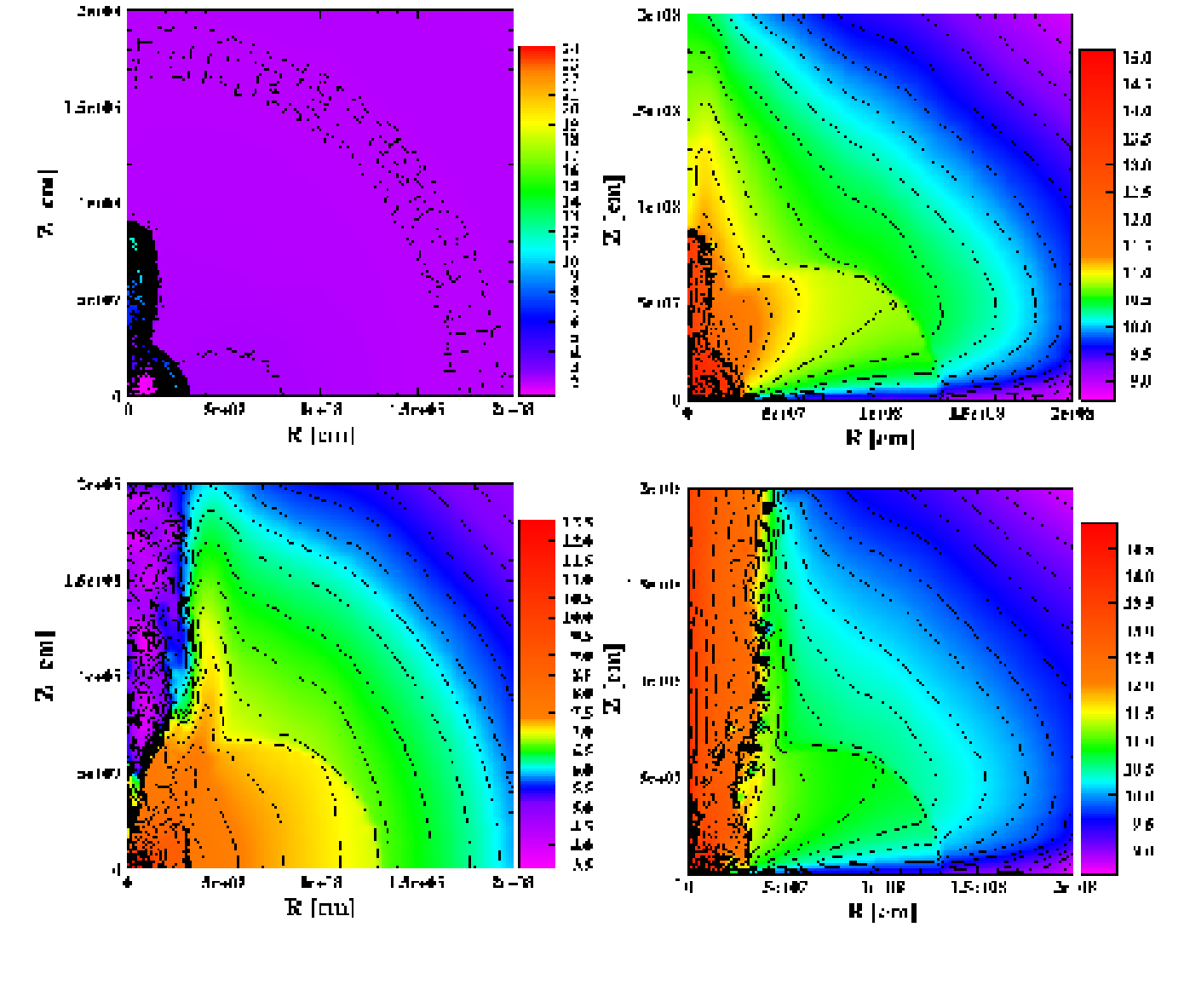}
\caption{
Various profiles of model B12TW4. 
Each figure shows 
entropy per nucleon at $t=32\ms$ (from core bounce) (left top), 
the logarithm of toroidal  magnetic field ($\gauss$) at $t=32\ms$ (right top),
the logarithm of density ($\gpcmc$) at $t=57\ms$ (left bottom) and 
the logarithm of toroidal magnetic field ($\gauss$) at $t=57\ms$ (right bottom).
The top figures show that  the magnetic field becomes strong behind the shock wave.
The bottom figures show the properties after propagation of the shock
 wave.
In the left figure,
It's found that a nozzle is formed near the rotational axis, where the
 density is significantly lowered.
}\label{fig:den_mag}
\end{figure}

We will first summarize some important magnetohydrodynamic
aspects which are common to our models.
The time evolution of our models can be specified by the three stages.
The first stage is between the onset of the core collapse and the core bounce.
The matter falls to the center of the core everywhere.
Therefore the magnetic fields frozen into the matter 
are compressed in this stage.
Next we call the epoch from the core bounce to
the propagation of the shock waves as the second stage. 
Since the 
rotation prevents the matter from infalling around the equatorial plane,
the matter near the rotational axis falls more quickly.
The core bounce first occurs near the axis.
High entropy regions are generated near the rotational axis,
and they begin to propagate along the axis.
The toroidal magnetic field is produced by the MHD processes in the shock wave.
The amplification of the magnetic fields is discussed elaborately in subsection 
\ref{sec:effects-magn-field}.
After the second stage,
 the shock wave is generated and propagates outward.
We call this stage as the third stage.
The properties of the shock waves are different
among the models, however 
it is common that
the toroidal magnetic field is transferred by the shock,
see the top panels of Figure \ref{fig:den_mag}.
Especially in the  models of the moderate magnetic field 
such as B10.5TW1, B10.5TW2 and B10.5TW4,
we found unique hydrodynamical features.
The shock wave tends to stall in these models.
After the stall the hydrodynamical features
resemble to those in the second stage.
The matter falls easily near the rotational axis, 
and the weaker shock wave is generated by the 
difference of the magnetic and the matter pressures 
between inside and outside of the shock wave.
After the launch of the shock the epoch goes to the third epoch.
The second stage and the third stage are repeated.
The arguments of specific behaviors among models are starting in the following.

\begin{figure}
\plotone{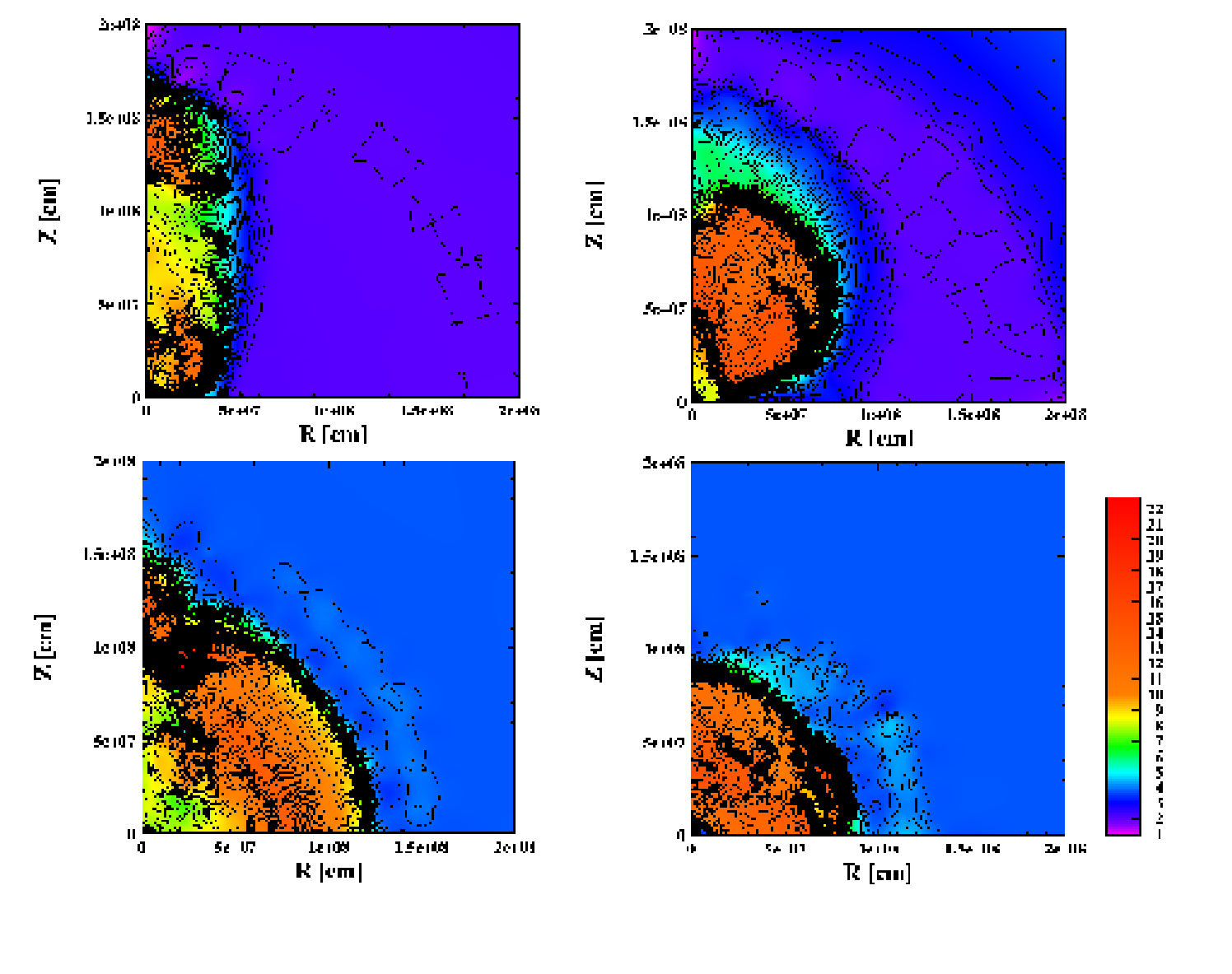}
\caption{
Profiles of the shock propagation in the various models.
Each panel demonstrates
 B12TW2 at $60\ms$ from core bounce (top left),
 B10.5TW1 at $127\ms$ (top right),
 B10.5TW2 at $219\ms$ (bottom left),
 B9TW4 at $404\ms$ (bottom right).
They show the  color coded contour plots of entropy ($k_{\mathrm{B}}$)
per nucleon.
Various profiles are found by changing the strength of the 
initial magnetic field and rotation.
}\label{fig:various}
\end{figure}

We show the differences of each model in the three stages
mentioned above.
Some of the final profiles are presented in Figure \ref{fig:various}.
It can be clearly seen in the figure that there is a
variety of final state of shocks.
We investigate the origin of this variety in the following.
Since the rotation does not affect qualitative features in our 
computation  so much,
we mainly focus on
the hydrodynamic features of each models caused by the initial
strength of the magnetic field.

In the strong magnetic models 
such as  B12TW1, B12TW2, and B12TW4, tightly collimated 
jet-like shock waves are found at the third stage.
Strong hoop stress is produced by the field wrapping at the launch of
the shock wave.
The launched shock waves are pushed by the strong magnetic pressure,
and propagate along the rotational axis,
while keeping its shape collimated.
The discussion for the collimation is given 
in the last of section \ref{sec:effects-magn-field}.
The matter around the polar axis
is expelled by the collimated jet-like shock waves.
These dynamic behaviors are seen in Figure \ref{fig:den_mag} in the case of B12TW4.
In all stages there are no qualitative differences  among these three models,
such as shape of the jet and the remaining configuration of the matter.
A major quantitative difference is the explosion energy discussed 
in section \ref{sec:explosion-energy}.

On the other hand, 
the shock waves become less collimated near the rotational axis
in the weak magnetic models such as  B9TW4, B9TW2, B9TW1, 
B0TW4, B0TW2 and B0TW1.
The prolate shock waves at the core bounce become rather spherical
after the shock stalls. 
This is caused by the magnetic and the matter pressure.
The difference among them is seen at a later stage rather.
In the case of the weak rotation,
the shock front revived by the magnetic and the matter pressure
propagates rather spherically.
On the other hand, in the case of the strong rotation  
the magnetic pressure produced by the field wrapping 
pushes the matter along the rotational axis at the launch of the shock. 
However this weak prolate shock wave becomes spherical soon due to 
weak hoop stress.

\begin{figure}
\plotone{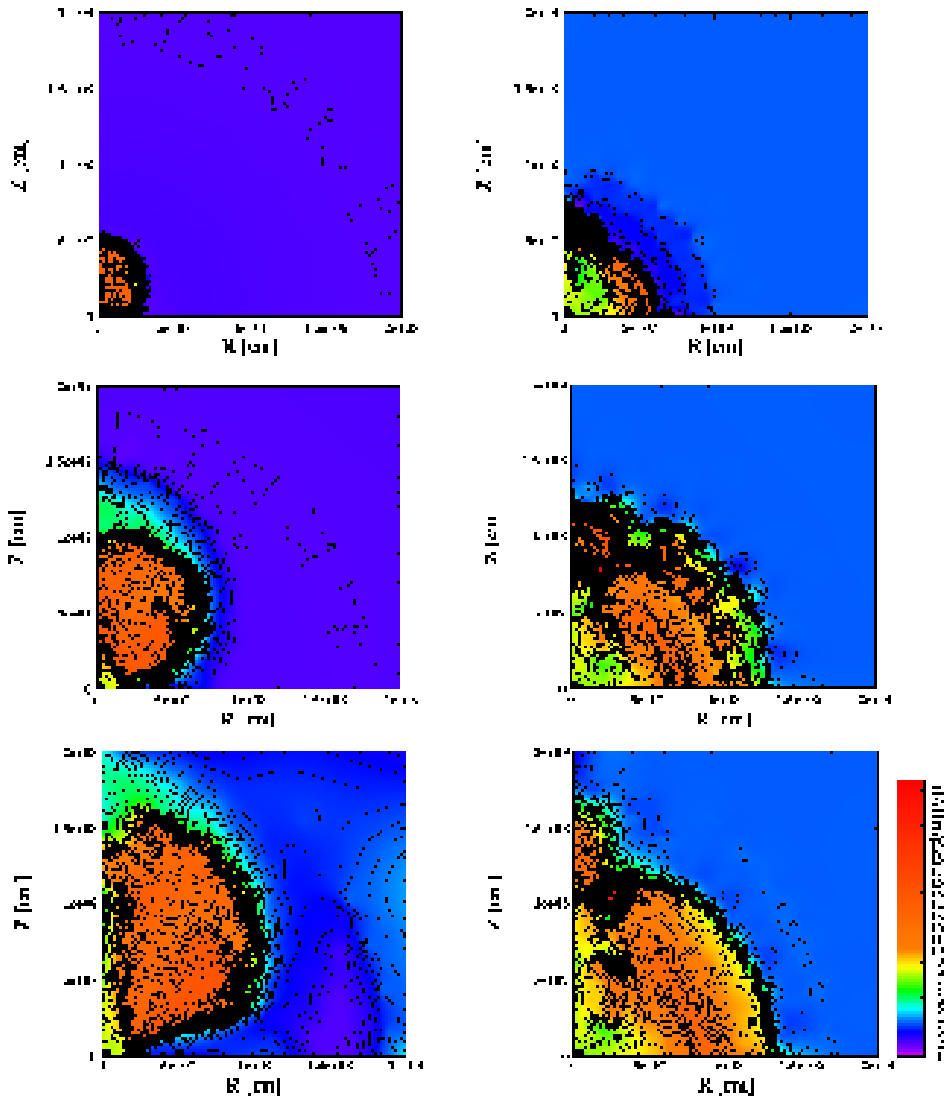}
\caption{
Time evolution of shock waves for B10.5TW1 and B10.5TW2.
These figures show color coded contour plots of entropy ($k_{\mathrm{B}}$)
per nucleon.
The three figures at the left side correspond to the model B10.5TW1,
at $t=52\ms$, $127\ms$, $219\ms$ from top to bottom, 
respectively.
The three figures at the right side correspond to the model B10.5TW2,
at $t=355\ms$, $405\ms$, $430\ms$ from top to bottom.
We discuss the features of this evolution in subsection \ref{sec:hydr-feat}.
}\label{fig:moderate_mag_profile}
\end{figure}

In the case of the moderate magnetic fields
such as B10.5TW4 and B10.5TW2,
the evolutions of the shock waves are  rather complicated,
see the right panels of Figure \ref{fig:moderate_mag_profile}.
In these cases,
jet-like shock waves are launched along polar axis,
and stall very soon.
$20\ms$ after the shock-stall,
the shock wave is found to be revived 
due to the magnetic and the matter pressure 
and propagates along the rotational axis again (right top).
However in this case, the hoop stress is not strong 
enough to keep the collimation of the prolate shock wave (right middle).
Then the prolate shock wave becomes spherical.
The shock front propagates repeating these behaviors (right bottom),
the shape of shock wave becomes spherical in the end. 
On the other hand, in the case of B10.5TW1 
(see the left panels of Figure \ref{fig:moderate_mag_profile}),
the stalled jet-like shock waves launched along polar axis 
are revived by the matter pressure (left top).
The weak rotation makes the second bounce robust as
the weak rotation makes first bounce robust.
The jet-like shock propagates along the rotational axis strongly (left middle).
After the shock propagates along rotational axis,
this robust shock wave becomes wider (left bottom).
The strength of the rotation rate makes these differences,
i.e. in the case of weak rotational model,
the strong shock wave is generated.
This shock front propagates before the matter behind the shock wave
expands.
In the rapid rotating model
the weak shock waves is likely to stall and
expand due to the matter and magnetic pressures.
For the analysis of these behaviors,
we focus on the value of plasma beta which is 
ratio of the gas pressure to the magnetic pressure
in the following section.

\subsection{Behavior of Magnetic Field and Degree of Collimation}\label{sec:effects-magn-field}
We explain how the magnetic fields affect
 the propagation of the shock waves in this subsection.
First we briefly explain two MHD processes so-called magnetorotational
instability (MRI) and field wrapping.
We focus on one of the MRI of the vertical component of the magnetic field,
because in our simulation initial magnetic field is vertical.
It is noted that 
this MRI produces the $X$ component of the magnetic field.
This MRI is important because 
the onset of the MRI requires the differential rotation of the core,
 i.e. $\del{X}{\Omega}<0$, which is globally content in the stellar collapse.
The time-scale of the maximum growth of the instability is evaluated as
$\tau_{\mathrm{\MRI}}=4\pi\left|X\pdel{X}{\Omega}\right|^{-1}$
\citep{balb98}.
If this value is constant,
the MRI process amplifies the $X$ component of the magnetic field
exponentially. 
As for the field wrapping,
it winds up poloidal magnetic field along axis,
and generate toroidal magnetic field.
This process also requires the differential rotation.
It's characteristic time scale is
$\tau_{\mathrm{\WRAP}}=4\pi\left|\frac{\pdel{t}{B_{\phi}}}{B_{\phi}}\right|^{-1}
=4\pi
\left|
\frac{B_{X}}{B_{\phi}}\left(X\pdel{X}{\Omega}\right)+
\frac{B_{Z}}{B_{\phi}}\left(X\pdel{Z}{\Omega}\right)
\right|^{-1}$
which is derived from induction equation
\citep{meir76}.
The definition of $\tau_{\mathrm{\WRAP}}$
shows that $\tau_{\mathrm{\WRAP}}$ becomes short if the poloidal magnetic 
field dominates toroidal component. 
It means that MRI can increase the seed poloidal component of magnetic field
which amplifies the toroidal magnetic field due to the field wrapping.
It should be noted that
since our simulation assumes axisymmetry,
the poloidal component can not be produced 
from the instability of the toroidal component.

\begin{figure}
\plotone{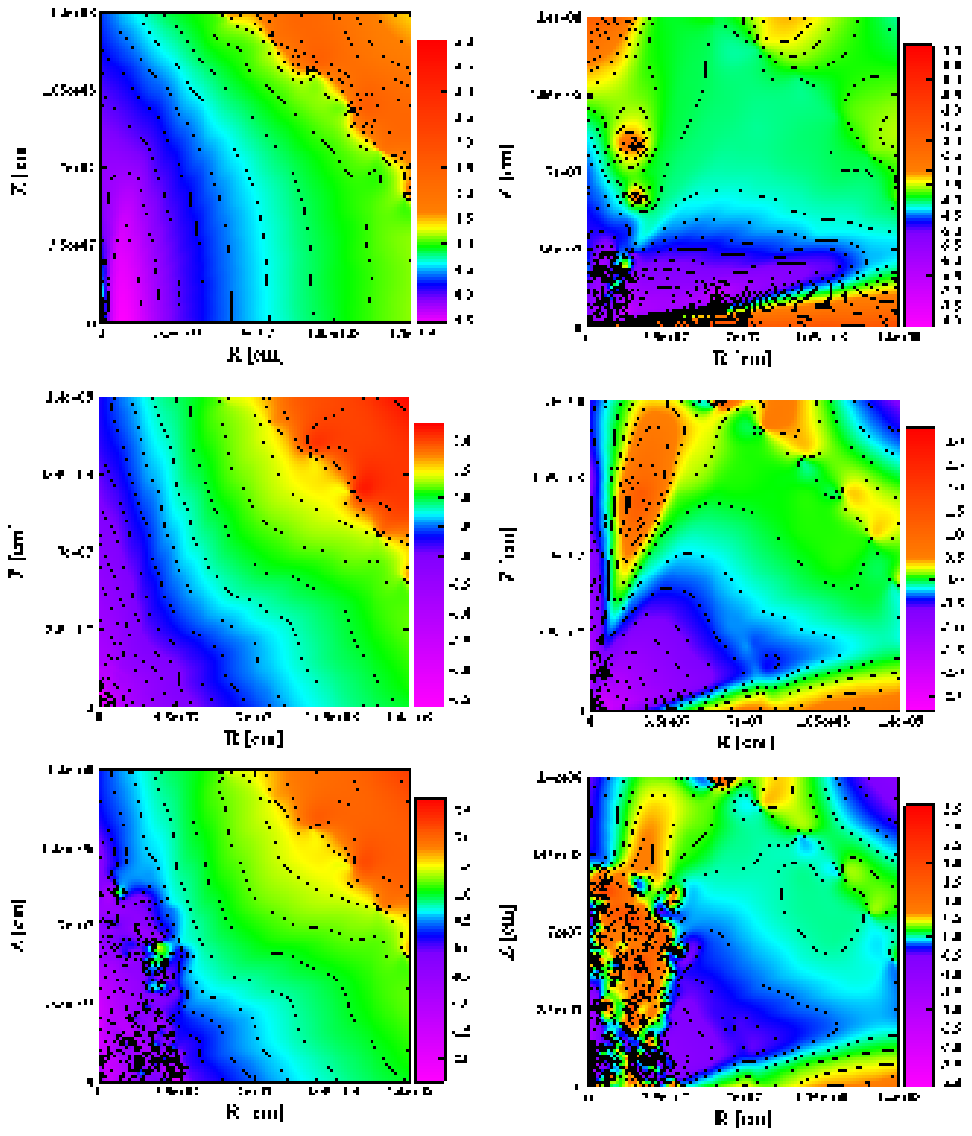}
\caption{The characteristic time scale of the maximum growth of the MRI and
the variation of the toroidal magnetic field.
This figure shows the contour of the logarithm of
the $\tau_{\MRI}[\mathrm{s}]$ (left) and 
the counter of the logarithm of the $\tau_{\WRAP}[\mathrm{s}]$
 (right) in the model of B12TW2.
The top figures show two time-scales at the onset of core
 collapse ($-172\ms$ from core bounce),
the middle figures  show these at near the core bounce ($5\ms$),
the bottom figures show these when jet propagates ($25\ms$).}
\label{fig:MRI_TIMESCALE}
\end{figure}

We show how magnetic fields are amplified in each stage
by comparing the growth time-scale of MRI and field wrapping
(see Figure \ref{fig:MRI_TIMESCALE}).
In the first stage,
the typical time-scale for this stage is less than $200\ms$ after the onset 
of the gravitational collapse.
Although the initial magnetic fields are parallel to the rotational axis,
the $X$ component is found to be formed by the infall of the matter.
See the left top panel of Figure \ref{fig:MRI_TIMESCALE},
the minimum values of $\tau_{\mathrm{MRI}}$
at this stage is about $300\ms$.
Since the time-scale is much longer than the dynamical time-scale,
thus the $Z$ component of magnetic field
is changed to the $X$ component mainly due to the
compression by the infalling material.
Since $\tau_{\mathrm{WRAP}}$ is very short,
the toroidal component of the magnetic field is rapidly amplified.
In the second stage,
we found that the toroidal magnetic field produced by the field wrapping
becomes to dominate over the poloidal component in the central region. 
Since the flow in this region is complicated 
by the convective flow,
the magnetic fields frozen into the flow become also  complicated.
These convections produce the $X$ component of the magnetic field 
from the initial $Z$ component.
In this stage, the MRI can grow in the center of core.
See the left middle panel of Figure \ref{fig:MRI_TIMESCALE}, 
the characteristic time scale 
of instability is about $ 10\ms$ at the center of the core.
Thus, MRI is likely to generate the $X$ component from the $Z$ component.
It is also seen the middle right panel in Figure \ref{fig:MRI_TIMESCALE} that
the time-scale of the wrapping is found to be shorter than 
that of the MRI. 
Therefore the toroidal magnetic field is predominantly generated
by the wrapping in this stage.
In the third stage,
since the toroidal magnetic field dominates over
the poloidal magnetic field behind the shock wave,
the wrapping time scale becomes very long.
In this era the main process of field amplification is the MRI.
This means that toroidal magnetic field is not dominantly generated 
by the field wrapping. 
This strong magnetic field is transferred by the propagation
of the shock wave.

\begin{figure}
\plotone{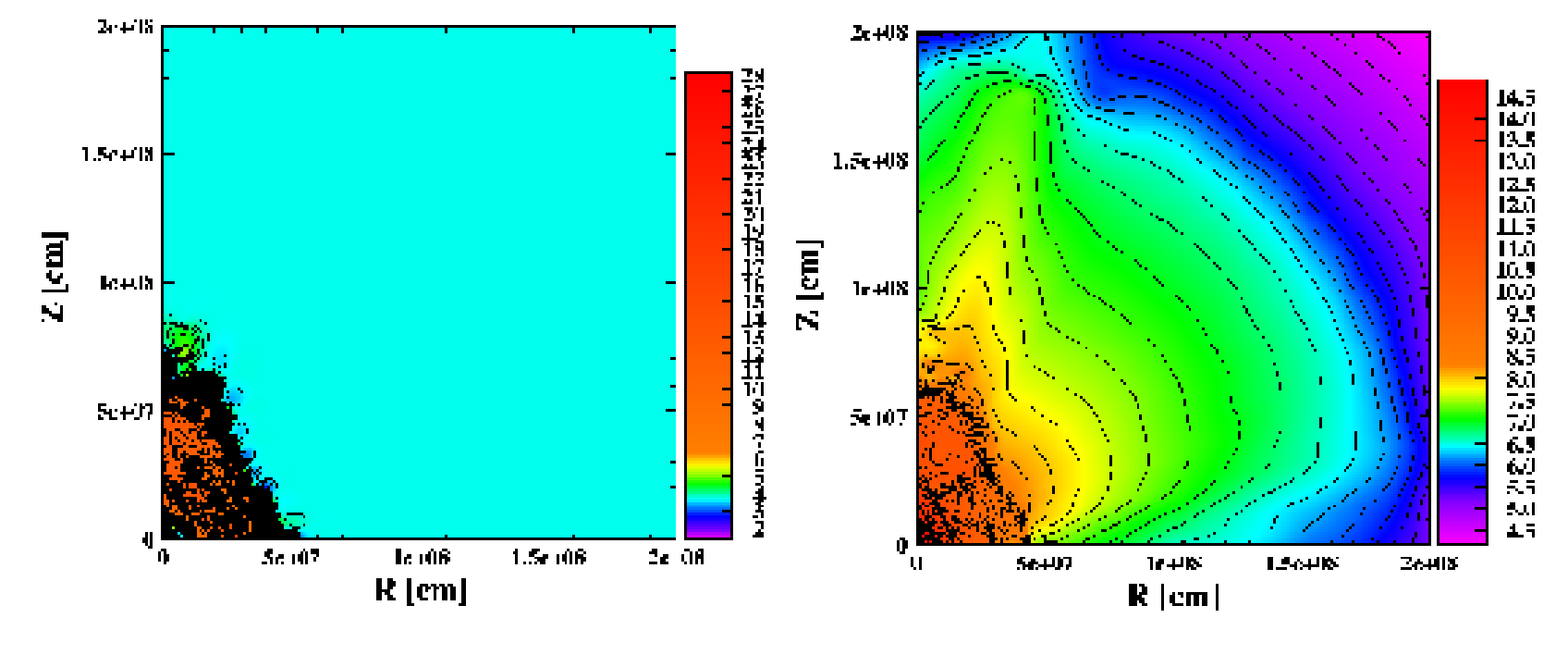}
\caption{
This figure shows 
entropy (left) and logarithm of the 
toroidal magnetic field (right) of model B9TW4.
Small jet is generated by the magnetic pressure.
The model B9TW4 begin with the weak magnetic field,
however we found the weak jet at the later stage ($366\ms$ from core bounce).
The toroidal magnetic field has been sufficiently amplified 
by the differential rotation in the later stage. 
}\label{fig:smalljet}
\end{figure}

Here it should be mentioned about the amplification of the magnetic field 
in the case of the weak initial magnetic field.
Even if initial magnetic field is weak,
long-time rotation increases strength of magnetic field.
In this era dominant process of field amplification is MRI.
We found the interesting example of these effects in our simulation.
The differential rotation amplifies strong toroidal magnetic field
even if the initial magnetic field is weak, see Figure \ref{fig:smalljet}.
Consequently the  small weak jet pushed by the magnetic pressure 
is generated as seen in Figure \ref{fig:smalljet}.
This situation occurs  in the only later stage ($366\ms$ from core bounce) 
because it takes long time to wind up the toroidal magnetic field 
if the initial poloidal magnetic field is weak.

\begin{figure}
\plotone{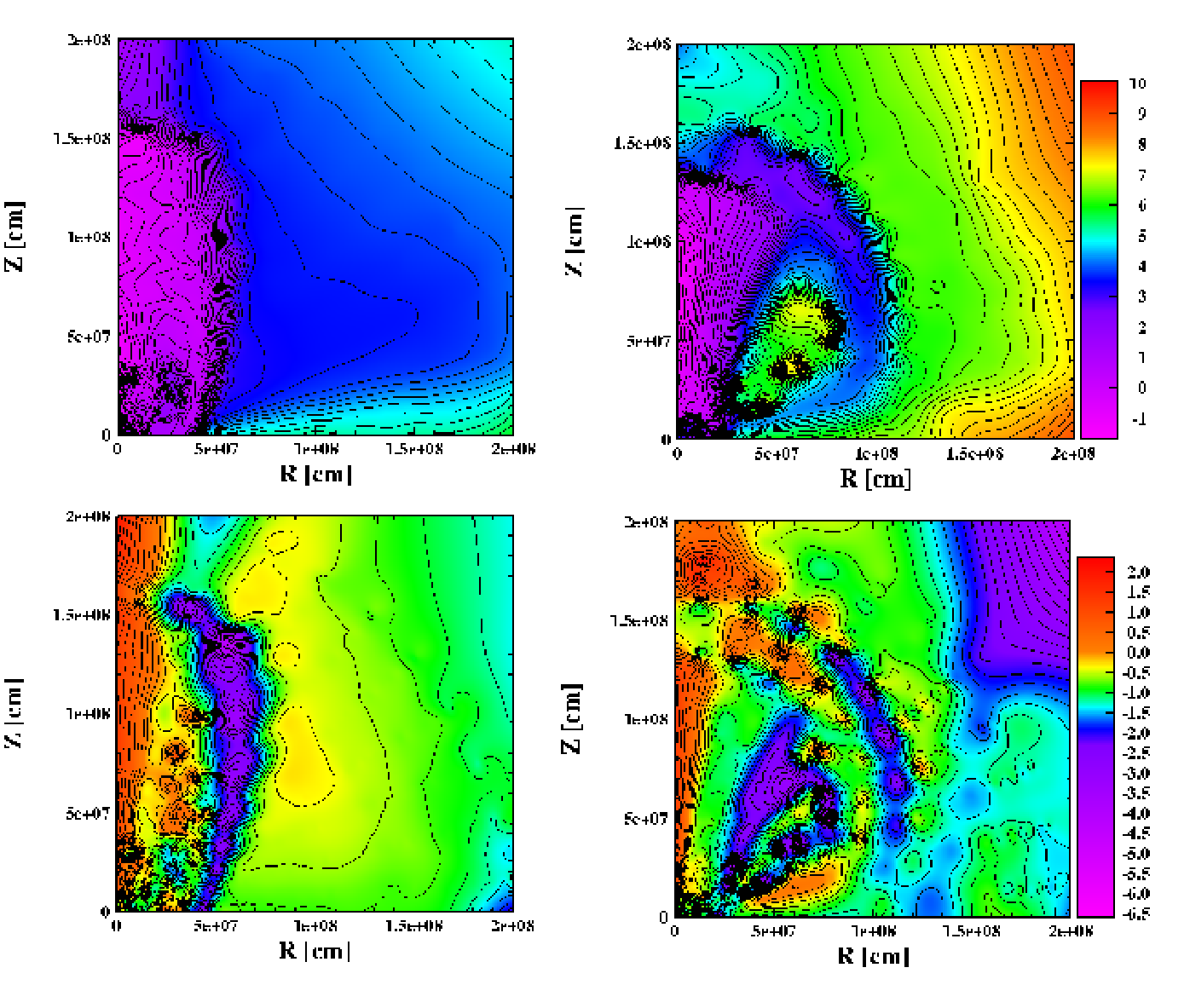}
\caption{
The top panel of the figure shows logarithm of the 
plasma beta of model B10.5TW1 (right, $243\ms$ from core bounce) 
and model B12TW2 (left, $60\ms$ from core bounce).
If $\beta<1$ inside the shock wave,
the jet is collimated.
On the other hand,
if there is region with $\beta>1$ inside the jet,
the shape of the jet becomes less collimated.
The bottom panel of figure the figure shows 
logarithm of 
the ratio of the hoop stress to the magnetic pressure.
This figure demonstrates clearly that
the shock wave is collimated by the hoop stress 
near the rotational axis.
}\label{fig:beta}
\end{figure}

Finally we explain the magnetic effect on the collimated shock wave.
First we show plasma $\beta$ which is the ratio of the magnetic pressure,
$P_{\vec{B}}=\frac{\vec{B}\cdot\vec{B}}{8\pi}$ to 
the gas pressure, $P$ in the top panel of Figure \ref{fig:beta}.
This figure shows the region where the magnetic field predominantly 
affects the dynamics of the matter. 
We can see jet is broaden if $\beta > 1$ in the interior of the shock waves.
However this is not enough to discuss the dynamical role of the magnetic
field, 
because the magnetic fields have two aspects for the collimation.
The hoop stress, $\frac{B_{\phi}^2}{X}$, collimate shock wave,
however the gradient of the magnetic pressure, $\frac{1}{2}\pdel{X}{B_{\phi}^2}$,
expand shock wave. 
Therefore we show the ratio of the hoop stress to 
the gradient of the magnetic pressure in the bottom panel of Figure \ref{fig:beta}.
This figure demonstrates that the hoop stress work near the rotational
axis and even if the case of the collimated shock wave magnetic pressure
is dominant at the shock front.
It means that even the collimated shock wave tend to expand parallel to
the $X$ axis.
Therefore the degree of the collimation is
 not determined by the magnetic field only.
The velocity of the shock front is another factor
which governs the degree of the collimation.
In the Figure \ref{fig:beta}, 
left panels are of B12TW2 and right panels are of B10.5TW1.
In B12TW2 the $Z$ component of the velocity of the shock front 
near the rotational axis at the launch is $4.0\times 10^9\cmps$ on the
contrary, that of B10.5TW1 is $1.5\times 10^9\cmps$.
This figure shows that 
the high $Z$ component of the velocity can make the shock front 
propagates before it expand to $X$ direction.
Such a strong collimated shock wave tends to
 propagate entirely through iron core.
On the other hand, the weak widely expanding jet or prolate shock wave
tends to stall.

\subsection{The Explosion Energy}
\label{sec:explosion-energy}
In this subsection,
we discuss the relation between the explosion energy
and the initial strength of the magnetic field
and the rotation.
We define the explosion energy as,
\begin{equation}
E_{\mathrm{explosion}}
=
\int_{\mathrm{D}}\d V\left( \frac{1}{2}\rho \vec{v}\cdot\vec{v}
-\rho\Phi+e+\frac{1}{8\pi}\vec{B}\cdot\vec{B}\right).\label{eq:2} 
\end{equation}
Integrating region, $\mathrm{D}$, represents  the domain where the integrand is 
positive.
This quantity is evaluated when 
shock wave  arrives at the radius of $1500\km$ at the $Z$ axis. 
We use this quantity as the measure of the strength of the explosion.

The explosion energy for each model is summarized in Table \ref{tab:expl}.
There is a general tendency that 
the rapid rotation decreases the explosion energy. 
The centrifugal force prevents the matter near the equatorial plane
 from falling rapidly,
therefore bounce occurs weakly. 
It is consistent to the previous work \citep{yama94,yama04}. 
Further more it's found that the explosion energy
monotonically increases with the initial magnetic field strength.
The magnetic pressure plays an important role in the explosion.
The strong toroidal magnetic field collimates the shock wave.
Thus it requires relatively low energy for shock wave to expel the matter 
near the rotational axis.
As a result,   
the strong magnetic field is favorable to the robust explosion
in the limited region.
On the other hand,
the weak magnetic fields
does not deviate the dynamics from pure rotation case
until the core bounce.
However it's found that they affect the dynamics in the later phase.
In fact, the model, which does not explode without magnetic fields,
explodes as a result of the field amplification.

To summarize, we find that the slow rotation and the 
strong magnetic field increase the explosion energy,
and that the even weak magnetic fields can contribute to the explosion 
at the later stage.

\section{Summary and Discussion}

We have done a series of two-dimensional magnetohydrodynamic simulations of the rotational
core collapse of a magnetized massive star. In this study, 
we have systematically investigated how the strong magnetic fields and the
rapid rotation affect the dynamics from the onset of the 
core-collapse to the shock propagation in the iron core.
By analyzing the properties of the shock waves and estimating the
explosion energies, we have obtained the results in the following.

\begin{enumerate}
 \item The explosion energy decreases with the initial rotation rates, 
       on the other hand, increases with the initial
       strength of the magnetic fields. As a result, it is found that 
       the combination of the weak rotation and the strong magnetic
       field prior to core collapse
       makes the explosion energy largest. 
 \item
       The toroidal magnetic fields amplified by the field wrapping 
       push the matter outwards as the magnetic pressure and collimate 
       the shock wave near the axis as the hoop stress. 
       On the other hand, the robust explosion near the rotational axis 
       makes the shapes of the shock waves prolate.
       Thus it is found that the degree of the collimation of the shock waves 
       is determined by the robustness of the asymmetric shock wave and 
       the competition between the hoop stress and 
       the magnetic and the matter pressure. 
     
 \item  In the models whose initial magnetic fields and initial rotation rates
       are large, the collimated shock waves are 
       found to carry the strong 
       magnetic fields in the central regions to the outer regions along
       the rotational axis. In addition, the regions with a relatively
       low density are found to appear like a nozzle along the rotational
       axis after the shock wave propagates out of the iron core. 

 \item 
       Even if the initial poloidal magnetic field is very weak, 
       the strength of the
       toroidal magnetic field increases with time as a result of 
       the field wrapping over a long rotation periods. As a result, 
       the dynamics in such models is found to be deviated
       significantly 
       from the pure rotation case.
\end{enumerate}

We should note that there are three major imperfections in our computations.
First of all, we did not include the neutrino heating. In addition to the
difficulty that hampers us to include it in the multidimensional
simulations, we intend
here to study the magnetohydrodynamic effects on the explosion mechanism 
as the first step. The combination of the neutrino heating and the
results obtained here will be mentioned in the forthcoming paper
\citep{kotakemhd, yamada04}.
Secondly, we assume that the initial magnetic field is just parallel to
the rotational axis. This assumption may not be so realistic
because the pulsars seem to require the dipole magnetic fields. 
We are also preparing to treat it as the next step.
Finally our simulation does not  allow all sorts of MRI.
Since our simulations assume axisymmetry that prohibits 
gradient of vector potential in the toroidal direction, the growth of 
the poloidal magnetic field due to MRI should be suppressed \citep{balb98}.
In order to investigate
how the MRI contributes to the amplification of the magnetic field, 
it is indispensable to perform the three dimensional MHD simulations
\citep{sano04,fryer04}.
In addition it is uncertain whether the number of the grid 
is enough to resolve the MRI of large wave number.
The dependence of the growth rate on the number of grid 
will be mentioned forthcoming paper. 

Bearing these caveats in mind,
we state some speculations based on our results.
As stated earlier, the models with the strong magnetic fields and the
weak rotation initially have the largest explosion energy.
It should be noted that the strength of 
the $Z$ component of the magnetic fields
in the protoneutron star is about $10^{15}\gauss$ in this model.
Thus such model might be related to
the formation of the so-called magnetars. It suggests that 
the explosion accompanied by the formation of the magnetars becomes
highly jet-like. 
However we obtain this result in the 
initially very strong magnetic field which may seem to be astrophysically 
unlikely.
It should be noted that more systematic parametric search is required.
We do not search the initial condition whose $|T/W|$ is less than $1\%$.
We expect that more slowly rotating core has more robust explosion
energy although too slow rotation which prevents the field
amplification,
should weaken the explosion energy.
We have to search more relevant initial condition for magnetar.  
It should be also mentioned that the regions of low density with the
strong toroidal magnetic fields appear along the rotational axis in the models.
Since the matter around equatorial plane still falls
slowly after the shock waves propagates out of the iron core, there 
might be another possibility that the highly magnetized protoneutron
star collapses to form a black hole in the later time. 
In the majority of the so-called collapsar models \citep{macf99,prog03},
the central black hole is assumed to be treated as an
initial condition, and then the nozzle near the rotational axis is
formed. The nozzle is considered to be the site for producing the gamma-ray bursts. On the other hand, it is expected in our models
that the nozzle is likely to be  formed
before the formation of a black hole. We think that the
model, in which the delayed
collapse of the protoneutron star to the black hole triggers the
gamma-ray bursts \citep{vietri}, should be also explored by the numerical simulations 
in context of core-collapse supernovae.

\acknowledgments

We are grateful to  S. Yamada for the fruitful discussions.
T. Takiwaki and K. Kotake are also grateful to S. Akiyama
for the discussion of MRI.
This study was partially supported
by Grants-in-Aid for the Scientific Research from the Ministry of Education, Science and
Culture of Japan through No.S 14102004, No. 14079202 and No. 16740134.

\begin{table}
\begin{center}
\caption{Models and Parameters }\label{tab:model}
\begin{tabular}{|c|ccc|}
\tableline\tableline
\backslashbox{$B_{0}(\mathrm{Gauss})$}{$T/|W|(\%)$} & 1\%
 &2\%  &4\%\\
\tableline
$0\gauss$          & B0TW1    & B0TW2    & B0TW4 \\
$10^{9}\gauss$     & B9TW1    & B9TW2    & B9TW4 \\
$10^{10.5}\gauss$  & B10.5TW1 & B10.5TW2 & B10.5TW4  \\
$10^{12}\gauss$    & B12TW1   & B12TW2   & B12TW4     \\
\tableline
\end{tabular}
\tablecomments{
This table shows the name of the models.
In the table they are labeled by the strength of the initial 
magnetic field and rotation.
$T/|W|$ represents the ratio of  the rotational energy to absolute value of
  the gravitational energy.
$B_{0}$ represents the strength of the poloidal magnetic field.}
\end{center}
\end{table}

\begin{table}
\begin{center}
\caption{Explosion Energy }\label{tab:expl}
\begin{tabular}{|c|ccc|}
\tableline\tableline
\backslashbox{$B_{0}(\mathrm{Gauss})$}{$T/|W|(\%)$} & 1\% &2\%  &4\%\\
\tableline
$0\gauss$          & 0  & 0  & 0  \\
$10^{9}\gauss$     & 0.05  & 0.00077 & 0  \\
$10^{10.5}\gauss$  & 4.7 & 0.73 & 0.03    \\
$10^{12}\gauss$    & 44  & 5.6  & 1.8     \\
\tableline
\end{tabular}
\tablecomments{
The explosion energy for each model is given.
For the definition of the explosion energy, see Eq. (\ref{eq:2}).
It's noted that they are normalized as $10^{50}$erg.}
\end{center}
\end{table}

\end{document}